\documentclass[aps, prc, superscriptaddress, twocolumn, floatfix, showpacs, altaffilletter, preprintnumbers, 10pt, notitlepage, nofootinbib]{revtex4-1} %
\usepackage[american]{babel}
\usepackage[T1]{fontenc}
\usepackage[utf8]{inputenc}
\usepackage{lmodern}
\usepackage{newtxtext,newtxmath} 
\usepackage{float}
\usepackage{bm}
\usepackage{mathtools}
\usepackage{amsmath} 
\usepackage{braket}
\usepackage{isotope} 
\usepackage{natbib}
\usepackage[squaren]{SIunits} 
\usepackage{booktabs} 
\usepackage{graphicx}
\usepackage{subfigure}
\usepackage{tabularx}
\usepackage{longtable}
\usepackage{titlesec}
\titleformat{\subsubsection}{\normalfont\bfseries\center\small}{}{0pt}{}
\titleformat{\section}{\normalfont\small\bfseries\center}{\thesection.}{0.8em}{}
\usepackage{multirow}
\usepackage{dcolumn} 
\usepackage{xcolor} 
\usepackage[colorlinks,hyperindex]{hyperref} 
\hypersetup
	{   colorlinks,%
		citecolor=blue,%
		linkcolor=blue,%
		urlcolor=blue,%
	}
\renewcommand{\arraystretch}{1.4}

\begin{document}
\title{\texorpdfstring{Lifetime measurements in neutron-rich odd-\textit{\textbf{A}} yttrium isotopes ($^{\textbf{93}-\textbf{99}}$Y): Investigation of shape coexistence and the intertwined quantum phase transition}{}}
\author{A.~Pfeil}
\email[Corresponding author: ]{apfeil@ikp.uni-koeln.de} \affiliation{Universität zu Köln, Institut für Kernphysik, 50937 Köln, Germany}
\author{N. Gavrielov}
\affiliation{Physics Department, Nuclear Research Center-Negev, P.O. Box 9001 Beer-Sheva, Israel}
\affiliation{School of Physics and Astronomy, Tel Aviv University, Tel Aviv, Israel}
\author{U. Köster}
\affiliation{Institut Laue-Langevin, 71 Avenue des Martyrs, 38042 Grenoble, France}
\author{Y. H. Kim}
\affiliation{Institut Laue-Langevin, 71 Avenue des Martyrs, 38042 Grenoble, France}
\author{N. Cieplicka-Ory\'{n}czak}
\affiliation{Institute of Nuclear Physics, PAN, 31-342 Krak\'{o}w, Poland}
\author{J. Dudouet}
\affiliation{Institut of Physique des 2 Infinis de Lyon, 4 Rue Enrico Fermi, 69100 Villeurbanne, France}
\author{A. Esmaylzadeh}
\affiliation{Universität zu Köln, Institut für Kernphysik, 50937 Köln, Germany}
\author{\L. W. Iskra}
\affiliation{Institute of Nuclear Physics, PAN, 31-342 Krak\'{o}w, Poland}
\affiliation{INFN Sezione di Milano, 20133 Milano, Italy}
\author{M. Ley}
\affiliation{Universität zu Köln, Institut für Kernphysik, 50937 Köln, Germany}
\author{J.-M. Régis}
\affiliation{Universität zu Köln, Institut für Kernphysik, 50937 Köln, Germany}
\author{D. Reygadas}
\affiliation{Institut Laue-Langevin, 71 Avenue des Martyrs, 38042 Grenoble, France}
\author{J. Jolie}
\affiliation{Universität zu Köln, Institut für Kernphysik, 50937 Köln, Germany}
\date{\today}
\begin{abstract}
\noindent Lifetimes of 16 excited states in the neutron-rich odd-$A$ nuclei $^{93-99}$Y were measured using fast-timing $\gamma$\nobreakdash-$\gamma$ coincidence spectroscopy with fast scintillation detectors at the LOHENGRIN recoil separator. Particular attention is given to the region around $N \approx 59$, where rapid changes in nuclear deformation and shape coexistence occur. The lifetimes, determined using the generalized centroid difference method, are compared with interacting boson\nobreakdash-fermion model calculations with configuration mixing, in which the odd-$A$ yttrium isotopes are described as a proton coupled to a bosonic core containing normal and intruder configurations. The results provide new constraints on theoretical descriptions of shape coexistence and structural evolution in neutron-rich nuclei near $A \approx 100$, particularly for odd-$A$ systems where experimental information remains limited.
\end{abstract}


\maketitle
\section{\label{Introduction}INTRODUCTION}
\noindent Nuclei in the neutron-rich region around $A \approx 100$, including the yttrium isotopes, exhibit a variety of shapes that may coexist within a single nucleus. Owing to their complex microscopic structure, these nuclei have attracted considerable theoretical interest. The nucleon-nucleon interaction plays a decisive role in shaping the shell structure and gives rise to diverse nuclear shapes, including quantum phase transitions associated with changes in deformation~\cite{cejnar2010quantum} and coexistence of different configurations~\cite{heyde2011shape}. A striking feature in this region is the rapid evolution of nuclear structure across the yttrium isotopic chain toward $N \approx 59$. Such abrupt changes are noteworthy, as nuclear properties typically evolve smoothly with increasing proton ($Z$) and neutron ($N$) numbers in partially filled shells. However, in nuclei with $Z \approx 39$ and $N \approx 59$, a sudden onset of deformation is observed. This behavior is evident in even-even strontium ($Z = 38$) and zirconium ($Z = 40$) isotopes, which provide benchmarks for the neighboring odd\nobreakdash-$A$ yttrium nuclei. In these systems, a pronounced drop in the first $2^+$ excitation energy, together with a strong increase in $B(E2; 0_1^+ \rightarrow 2^+_1)$ values~\cite{pritychenko2016tables}, signals the onset of deformation. Additional evidence comes from the nonlinear increase of charge radii along the rubidium, strontium, yttrium, zirconium, and niobium chains~\cite{angeli2013table}. In particular, yttrium isotopes offer a unique probe of this region due to the presence of an unpaired proton, enabling direct access to the interplay between single-particle and collective degrees of freedom. Unlike even-even systems, they allow spectroscopic studies of specific proton configurations coupled to a rapidly evolving core. The onset of deformation near $N \approx 59$ was first interpreted by Sheline et al. as arising from the interplay of spherical and deformed configurations, including intruder states, leading to shape coexistence~\cite{sheline1972shell}. Microscopically, this behavior is linked to occupation of $\pi g_{9/2}$ orbitals, filling of $\nu h_{11/2}$ neutron orbitals, and rearrangement of lower-lying neutron configurations. Proton subshell closures at $Z = 38$ ($\pi p_{3/2}$) and $Z = 40$ ($\pi p_{1/2}$), together with neutron subshell closures at $N = 50$ ($\nu g_{9/2}$), $56$ ($\nu d_{5/2}$), and $58$ ($\nu s_{1/2}$), combined with weak residual proton-neutron interaction, lead to semi-magic-like behavior for $50 \leq N \leq 58$ in strontium, yttrium, and zirconium isotopes. At $N = 60$, a strongly prolate-deformed rotational character becomes favored. In contrast, deformation evolves more gradually in the Mo$-$Pd region across $N = 58-60$, where triaxial shapes are often observed~\cite{rodriguez2010charge}. At lower proton numbers, nuclei such as $^{96,98,100}$Kr show reduced collectivity for $N \geqslant 60$, attributed to reduced $\pi g_{9/2}$ occupation and shape coexistence~\cite{PhysRevLett.118.242501}. The rapid structural change in this region remains a challenge for theory. A common interpretation invokes competition between a near\nobreakdash-spherical shell-driven configuration and a deformed intruder configuration involving multi-particle-multi-hole excitations, governed by the proton\nobreakdash-neutron interaction~\cite{sheline1972shell}. Odd-$A$ yttrium isotopes are particularly sensitive probes, as the valence proton strongly couples to both configurations, making them well suited for studying configuration mixing and deformation onset. The complexity of low-energy spectra makes these nuclei important benchmarks for structure models. Extensive studies of neighboring even-even systems provide essential reference points. These include evidence for shape coexistence in $^{96}$Zr~\cite{PhysRevC.99.031304} and $^{98}$Zr~\cite{PhysRevC.102.064314}, a quantum phase transition at $N = 60$~\cite{PhysRevC.96.054323}, and $\gamma$-soft or triaxial structures in $^{100,102}$Zr~\cite{PhysRevC.100.014319}, supported by theory~\cite{Nomura2020,PhysRevC.99.064324}. In contrast, systematic information on odd-$A$ yttrium isotopes remains limited, despite their importance for single-particle-collective coupling in this region. Evidence for coexistence of spherical and deformed configurations has been reported in $^{96-98}$Sr~\cite{PhysRevC.41.1115,PhysRevLett.116.022701}, $^{96-104}$Y~\cite{PhysRevC.41.1141,Iskra_2017,CHEAL2007133}, and $^{98-100}$Zr~\cite{PhysRevC.49.1379,PhysRevC.98.041302,URBAN2001605,urban2003first,PhysRevLett.117.172503}. Strong $\pi g_{9/2}$-$\nu g_{7/2}$ spin-orbit interactions weaken the $N = 56, 58$ and $Z = 38, 40$ subshell closures, enhancing collectivity~\cite{federman1979unified}. Studying yttrium at the transitional boundary near $N \approx 59$ provides direct insight into the competition between single-particle motion and deformation. Despite extensive work, no unified theoretical description reproduces all features of shape coexistence in this region~\cite{heyde2011shape}. Information on collectivity can be obtained from electromagnetic transition rates in odd-$A$ yttrium nuclei. The reduced $E2$ transition probability $B(E2)$ reflects quadrupole collectivity and is extracted from lifetimes of stretched intraband $E2$ transitions. In addition, intraband $B(M1)$ values provide sensitivity to single-particle configurations. \\
The structure of this work is as follows. Section~\ref{Experiment} describes the experimental setup. The fast-timing method is outlined in Section~\ref{Fast-timing}. Analysis procedures are presented in Section~\ref{Analysis}. Sections~\ref{sec:framework}$-$\ref{discuss} present theoretical calculations and comparisons with experiment. Conclusions are given in Section~\ref{sec:conclusion}.
\section{EXPERIMENTAL DETAILS}
\label{Experiment}
\noindent The yttrium isotopes $^{93}$Y, $^{95}$Y, $^{97}$Y, and $^{99}$Y were investigated in two separate experimental campaigns, each utilizing slightly different setups. The first campaign comprised two experiments involving $A=95$ and $A=97$ isobars. In a second campaign, two experiments were performed to measure lifetimes in the $A=93$ and $A=99$ isobars. In the following, the experimental configuration used for the measurements of the odd-$A$ isobars, including $^{93-99}$Y, is described. \\
Fission fragments were generated by neutron-induced fission of $^{235}$U oxide targets (7\,cm long, 0.6 or 0.7\,cm wide and thicknesses ranged from 119 to 363\,$\mu \text{g}/\text{cm}^2$) placed in a neutron flux of approximately $\approx 5 \times 10^{14} \, \text{cm}^{-2}\text{s}^{-1}$ in the H9 beam tube of the high-flux reactor operated by Institut Laue-Langevin, Grenoble. All fission targets were covered with a 0.25\,$\mu$m thick nickel foil to reduce sputtering~\cite{koster2010experience}. \\
Recoiling fragments were separated by the LOHENGRIN recoil separator~\cite{ARMBRUSTER1976213,fioni1993reduction} according to their mass-over-ionic charge ratio. $A = 95$ was separated at $q = 20$, the other masses at $q = 21$. Kinetic energies between $E = 92$\,MeV and 100\,MeV were selected to optimize intensity and purity of the mass-separated beams. The ions traversed an ionization chamber (IC) filled with 15\,mbar isobutane, before being implanted into an aluminum foil mounted in the center of the detection setup. The detection system consisted of four cerium-doped lanthanum bromide (LaBr$_3$(Ce)) scintillation detectors, each with dimensions of 1.5” $\times$ 1.5”. These detectors were arranged as close as possible around the ionization chamber to optimize efficiency, following the layout of previous studies~\cite{PhysRevC.100.064309}. This configuration yielded a detector\nobreakdash-to\nobreakdash-implantation\nobreakdash-zone distance of 2.2\,cm. The implantation zone itself was a rectangular area measuring approximately $1 \times 3$\,cm$^2$~\cite{PhysRevC.100.064309}. The photopeak efficiency of each LaBr$_3$(Ce) detector was about 3.7\% at 122\,keV~\cite{REGIS2020163258}. Complementing the scintillator setup, high\nobreakdash-purity germanium (HPGe) clover detectors (two in the first campaign and one in the second), each comprising four individual crystals, were installed directly beneath the implantation site. This provided high-resolution $\gamma$-ray spectroscopy with a combined photopeak efficiency of about 3.8\% at 122\,keV~\cite{REGIS2020163258}. \\
The $A/q$ separation produced a cocktail beam containing the isobars of interest together with a small (few percent) admixture of contaminants having similar $A/q$ ratios. For the $A=95$ setting, contaminants with masses $A=90$ and 100 were observed, while masses $A=88$, 92, and 102 were identified in measurements of the $A=97$ isobars. In addition, contaminants corresponding to mass $A=88$ were observed for the $A=93$ setting, whereas mass $A=104$ contaminated the $A=99$ setting. \\
The flight time of ions from the target to the focal plane is about 1.7\,$\mu$s. Consequently, the present experiment provided access to $\gamma$-ray cascades populated either through isomeric transitions (IT), including those from the $9/2^+_1$ state in $^{95}$Y (T$_{1/2} = 52.6(12)\,\mu$s~\cite{PhysRevC.79.044304}) and the ($17/2^+_2$) state in $^{99}$Y (T$_{1/2} = 8.6(8)\,\mu$s~\cite{MEYER1985510}), or populated by $\beta^-$ decay of strontium isotopes to their corresponding yttrium daughters, respectively. Here, the relative $\beta$-decay activity of a specific isobar roughly scales with its cumulative fission yield. In $^{235}$U(n$_{\text{th}}$,f) these are 6.3\% for $^{93}$Sr, 5.4\% for $^{95}$Sr, 1.9\% for $^{97}$Sr, then dropping to 0.2\% for $^{99}$Sr~\cite{JEFF4.0}. Lifetime measurements in $^{99}$Y were performed only via IT decay of the ($17/2_2^+$) microsecond isomer while in $^{95}$Y both, $\beta$ decay from $^{95}$Sr and IT decay of the $9/2^+_1$ microsecond isomer were used. While the independent fission yields of the yttrium isotopes are known from databases (1.1\% for $^{95}$Y and 1.9\% for $^{99}$Y~\cite{JEFF4.0}), the isomeric ratio of the microsecond isomers, i.e. their relative population, is not available from databases and has been determined in the present experiment. For more detailed information about the fission yields of $A = 97,99$ isobars, the reader is referred to Refs.~\cite{PhysRevC.100.064309,PhysRevC.108.034310}. All data recorded during the experiments can be found in Refs.~\cite{ILLDataA93_95,ILLDataA97,ILLDataA99}.
\section{FAST-TIMING METHOD}
\label{Fast-timing}
\begin{figure}[t!]
	\centering
	\includegraphics[width=8.6 cm]{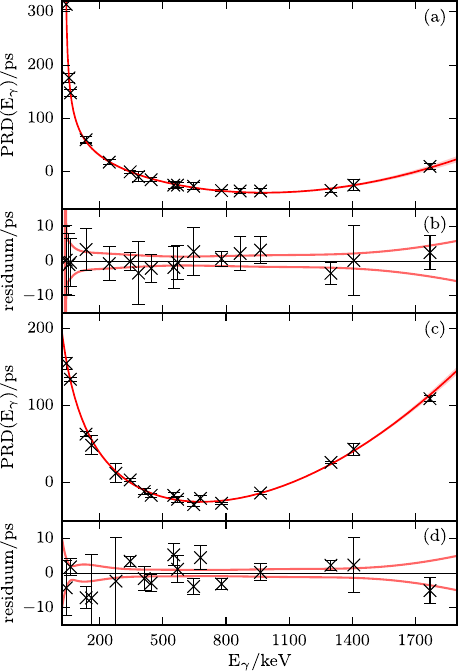}
	\caption{Resulting $PRD$ curves for the setup measuring $^{95}$Y and $^{97}$Y (a) and for the setup measuring $^{93}$Y and $^{99}$Y (c), determined using $^{133}$Ba, $^{152}$Eu, $^{185}$Os, $^{187}$W, and $^{207}$Bi calibration sources, with Eq.~(\ref{PRD}) as fit function. Known lifetimes for the data points are taken from Refs.~\cite{KHAZOV2011855,MARTIN20131497,knafla2023improving,WU2005619,BASUNIA2009999,KONDEV2011707}. (b,d) Fit residuals of the corresponding $PRD$ curves. The 1$\sigma$ uncertainty band is plotted in red.}
	\label{TW}
\end{figure}

\noindent To measure the lifetimes of excited states in odd-$A$ yttrium isotopes ($^{93-99}$Y), the $\gamma$-$\gamma$ fast-timing technique was employed using the LaBr$_3$(Ce) scintillation detectors. This method relies on determining the time interval between two coincident $\gamma$ rays. The experimental setup typically includes at least two detectors, constant fraction discriminators (CFDs), a time\nobreakdash-to\nobreakdash-amplitude converter (TAC), and a data acquisition system (DAQ), all of which are essential for capturing both time and energy signals. \\
The measured $\gamma$-$\gamma$ time difference spectra feature two independent distributions, known as delayed and anti\nobreakdash-delayed. Assuming no background interference, the delayed distribution $D(t)$ is obtained by convolution of the prompt response function (PRF), denoted as $P(t' - t_0)$, with an exponential decay law~\cite{REGIS201083}:
\begin{align}
D(t) = n\lambda \int_{-\infty}^{t} P(t'-t_0) e^{-\lambda(t - t')} dt' + n_r; \quad \lambda = \frac{1}{\tau},
\label{D(t)}
\end{align}
\noindent where $n$ is the total number of counts in the distribution and $n_r$ represents constant random (background) counts. The centroid $C^D$ of the delayed time spectrum, representing its mean value, is defined by~\cite{REGIS201672}:
\begin{align}
C^D = \langle t \rangle = \frac{\int_{-\infty}^{\infty} t \, D(t) dt}{\int_{-\infty}^{\infty} D(t) dt},
\label{C^D}
\end{align}
\noindent with an associated statistical uncertainty of approximately $\delta C^D \approx \delta t = \sigma / \sqrt{n}$. As detailed in Ref.~\cite{Bay}, the centroid of the delayed distribution is shifted by the mean lifetime relative to the centroid $t_0$ of the PRF. \\
Using the generalized centroid difference (GCD) method, level lifetimes can be extracted from the following relation~\cite{REGIS201083,REGIS2013191}:
\begin{align}
\Delta C(E_{\text{feeder}}, E_{\text{decay}}) = PRD(E_{\text{feeder}}, E_{\text{decay}}) + 2\tau.
\label{deltaC}
\end{align}
\noindent Here, $\Delta C$ is the difference in centroids between the delayed and anti-delayed spectra and $PRD$ is the prompt response difference—a quantity that captures the energy-dependent timing characteristics of the detector setup~\cite{REGIS2013191}. Accurate calibration of the $PRD$ is critical and was carried out using full-energy peaks (FEPs) from the calibration sources $^{133}$Ba, $^{152}$Eu, $^{185}$Os, $^{187}$W, and $^{207}$Bi (Note that not all possible cascades of the calibration sources were accessible in the different campaigns leading to slightly different data points used for the fitted curves). A detailed procedure is outlined in Ref.~\cite{REGIS201672}. The resulting $PRD$ curves for both campaigns are shown in Fig.~\ref{TW} and were obtained by fitting the following empirical function~\cite{REGIS2014210}:
\begin{align}
PRD(E_{\gamma}) = \frac{a}{\sqrt{E_{\gamma} + b}} + cE_{\gamma}^2 + dE_{\gamma} + e.
\label{PRD}
\end{align}
\noindent To determine $PRD(E_{\text{feeder}}, E_{\text{decay}})$, the values of $PRD$ at the feeder and decay transition energies are simply subtracted. \\
It is important to note that the validity of Eqs.~(\ref{D(t)}) and~(\ref{deltaC}) relies on the absence of time-correlated background. To account for such background, a correction can be applied by interpolating timing data from several off-peak background gates. An effective approach to compensate for this influence is to introduce a correction factor $\tilde{t}_{\text{cor}}$, as outlined in Refs.~\cite{PhysRevC.96.054323,PhysRevC.99.024326}:
\begin{align}
\Delta C = \Delta C_{\text{exp}} + \tilde{t}_{\text{cor}}.
\end{align}
\noindent Both the feeding and decaying transitions are affected by background contributions. To correct for these, a weighted average of individual correction terms is used~\cite{REGIS2020163258}:
\begin{align}
\tilde{t}_{\text{cor}} = \frac{P/B(E_f) \, t_{\text{cor}}(E_i) + P/B(E_i) \, t_{\text{cor}}(E_f)}{P/B(E_i) + P/B(E_f)},
\label{t_cor tilde}
\end{align}
\noindent where $P/B$ denotes the peak-to-background ratio and $t_{\text{cor}}(E_i)$ is calculated as:
\begin{align}
t_{\text{cor}}(E_i) = \frac{\Delta C_{\text{exp}} - \Delta C_{\text{Compton}}}{P/B(E_i)}.
\label{t_cor}
\end{align}
\noindent In this expression, $\Delta C_{\text{Compton}}$ is the interpolated centroid difference due to Compton background underneath the FEPs. All associated uncertainties are calculated using standard Gaussian error propagation.
\section{ANALYSIS}
\label{Analysis}
\begin{figure*}[h]
	\centering
	\includegraphics[width=16.3 cm]{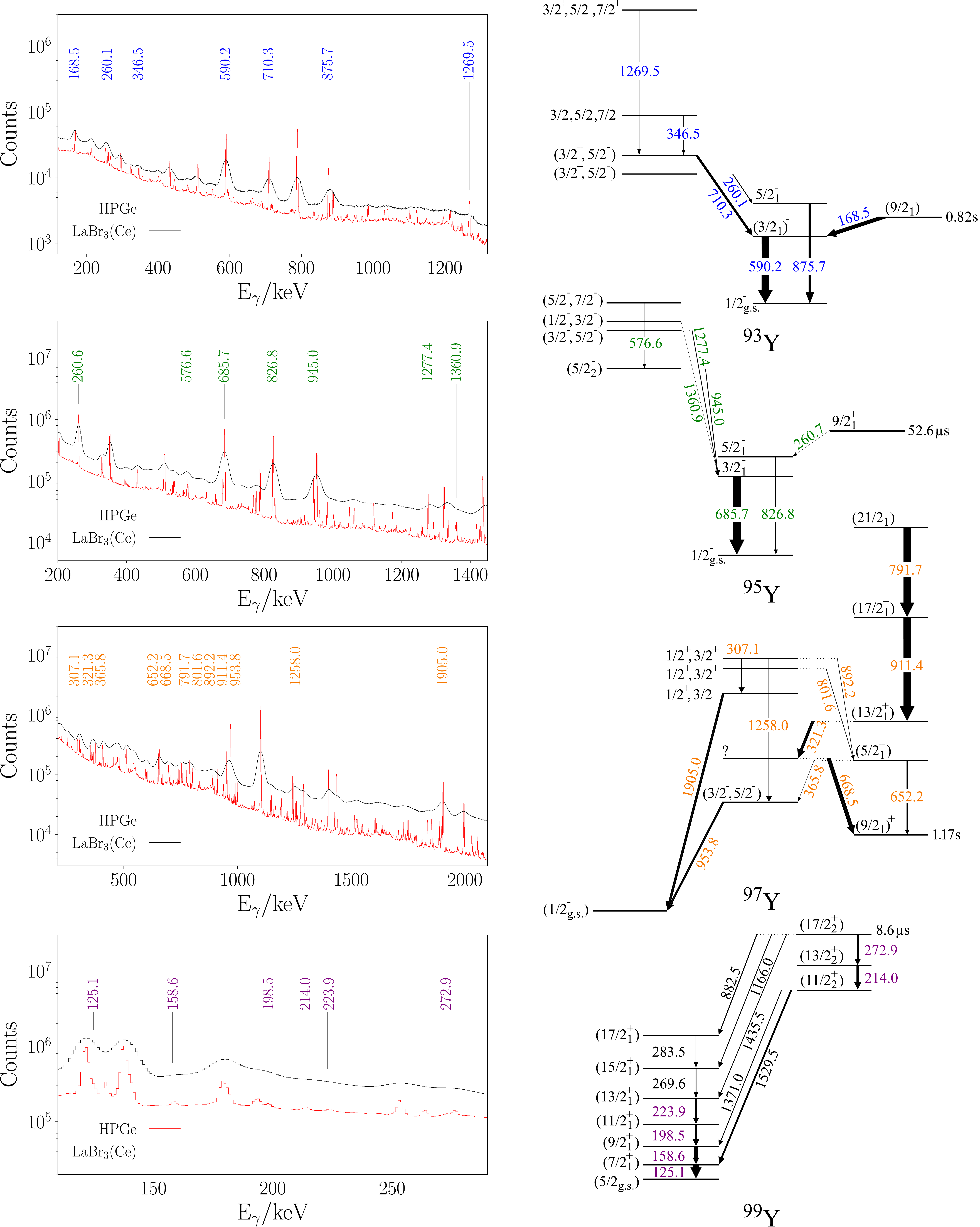}
	\caption{On the left side, full projections of LaBr$_3$(Ce)$-$LaBr$_3$(Ce) (black) and LaBr$_3$(Ce)$-$HPGe (red) twofold coincidences without any energy conditions applied and within a 1 $\mu$s coincidence window are shown. Relevant transitions~\cite{BAGLIN20111163,BASU20102555,NICA2010525,Browne2011} for the determination of lifetimes in the odd-$A$ $^{93-99}$Y nuclei are marked in blue, green, orange, and purple, respectively. The peaks that are not marked mostly stem from isobars populated after $\beta^-$ decay and masses which have a similar $A/q$ ratio (for a detailed description see Sec.~\ref{Experiment}). On the right side, the corresponding partial energy level scheme after IT and $\beta$ decay are displayed. The labeled isomeric $9/2_1^+$ states in $^{93,95,97}$Y and the $17/2_2^+$ state in $^{99}$Y are annotated with their half-lives. Known band structure, such as yrast and positive-parity bands, are distinguished from states with as yet unknown spin$/$parity assignments. \label{full_proj}}
\end{figure*}
\noindent In this work, lifetimes of 16 states in the odd-$A$ yttrium isotopes $^{93}$Y, $^{95}$Y, $^{97}$Y, and $^{99}$Y were determined. For $^{93}$Y, the first $(3/2)^-$ and $5/2^-$ states were measured, together with the $(3/2^+, 5/2^-)$ level at 1300.5\,keV, for which the spin$/$parity assignment remains tentative. The study of $^{95}$Y yielded lifetimes for the first $3/2^-$ and $5/2^-$ states, as well as for the second $(5/2^-)$ state. In the case of $^{97}$Y, three levels with uncertain spin$/$parity—[$(3/2^-, 5/2^-)$ at 953.8\,keV, "?" at 1336.0\,keV, and $1/2^+, 3/2^+$ at 1904.9\,keV]—were investigated in addition to the first $(5/2^+)$, $(13/2^+)$, and $(17/2^+)$ states. In $^{99}$Y, lifetimes for the first $(7/2^+)$, $(9/2^+)$, and $(11/2^+)$ states and the second $(13/2^+)$ state were obtained. Note that the lifetime value obtained for the $(13/2^+_2)$ state was not adopted because of slight contamination of the 269.6\,keV transition. The result is, however, consistent with the literature value within 1\,$\sigma$. The left side of Fig.~\ref{full_proj} shows the full projection spectra of LaBr$_3$(Ce) and HPGe twofold coincidences in the relevant energy range. Only $\gamma$ rays detected within a 1\,$\mu$s coincidence window—either between two LaBr$_3$(Ce) detectors or between a LaBr$_3$(Ce) and an HPGe detector—are included. The superior energy resolution of the HPGe detectors is used to validate the LaBr$_3$(Ce) spectra and to identify contaminant lines. In addition, for states whose feeding and$/$or decay energies are affected by coincident high-energy $\gamma$-ray feeding, the contribution of the Compton background beneath the peak of interest was evaluated by gating in the background on both sides. In all cases, it was found to be negligible and was also accounted for by the time-correlated background correction. Unmarked peaks predominantly arise from isobars populated after $\beta$ decay and from masses $A = 88, 90, 92, 100, 102$, and 104, which have similar $A/q$ ratios depending on the selected mass of interest (see Sec.~\ref{Experiment}). The right side of Fig.~\ref{full_proj} displays the corresponding partial level schemes following IT and $\beta$ decay, including all transitions relevant for the lifetime analysis. Most lifetimes were extracted from twofold [LaBr$_3$(Ce)$-$LaBr$_3$(Ce)] coincidences. For the states in $^{99}$Y, the available statistics allowed the use of threefold [HPGe$-$LaBr$_3$(Ce)$-$LaBr$_3$(Ce)] coincidences by summing matrices with different HPGe gates, yielding improved selectivity for the cascades of interest. The $5/2_1^-$ state in $^{95}$Y constitutes a special case, for which threefold [IC$-$LaBr$_3$(Ce)$-$LaBr$_3$(Ce)] coincidences were employed (see Sec.~\ref{IC_state}). All lifetimes were determined using the GCD method. Gate widths between 6 and 20\,keV were used, depending on the energy of the FEPs, the statistics, and the position of nearby peaks. In the following, the procedure for determining lifetimes is illustrated for the $5/2_1^-$ state in $^{93}$Y; the same approach was applied to all other states. Table~\ref{tab:alldata} summarizes all relevant information for the state of interest, including the cascades used, peak-to-background ratios, and the final adopted values. 
\subsection{\texorpdfstring{$\textbf{5}/\textbf{2}_{\textbf{1}}^-$ state in $^{\textbf{93}}$Y}{}}
\noindent The first $5/2^-$ state in $^{93}$Y is located at an excitation energy of 875.9\,keV. Its lifetime was determined using the $(3/2^+, 5/2^-) \rightarrow 5/2_1^- \rightarrow 1/2^-_{\text{g.s.}}$ (260.1\,keV$-$875.7\,keV) cascade. The corresponding gated energy spectra are shown in Figs.~\ref{Results}~(a,b). The peak-to-background ratios are approximately 2.2 for the 260.1\,keV transition and 3.3 for the 875.7\,keV transition. Both peaks in the LaBr$_3$(Ce) spectra are free of observable contamination. The background correction, including interpolation of the time-correlated Compton background at the energies of both FEPs, is presented in Figs.~\ref{Results}~(c,d). The green bar indicates the width of the applied gates. The background was constructed using 20\,keV gates stepped in 1\,keV intervals to ensure sufficient statistical quality for describing the local background behavior. To avoid correlations between adjacent data points, only points separated by at least 20 steps were included in the fit used to interpolate the Compton background beneath the peaks of interest. These data points are plotted in red and their positions are marked as red arrows in the energy spectra for better clarity. Figure~\ref{Results}~(e) shows the resulting (anti-)delayed time distributions obtained from the analysis using the GCD method. This leads to a final adopted lifetime value of $\tau_{{5/2}_1^-} = 16(5)$\,ps.
\subsection{\texorpdfstring{$\textbf{5}/\textbf{2}_{\textbf{1}}^-$ state in $^{\textbf{95}}$Y}{}}
\label{IC_state}
\begin{figure*}[t!]
	\centering
	\includegraphics[width=12.7 cm]{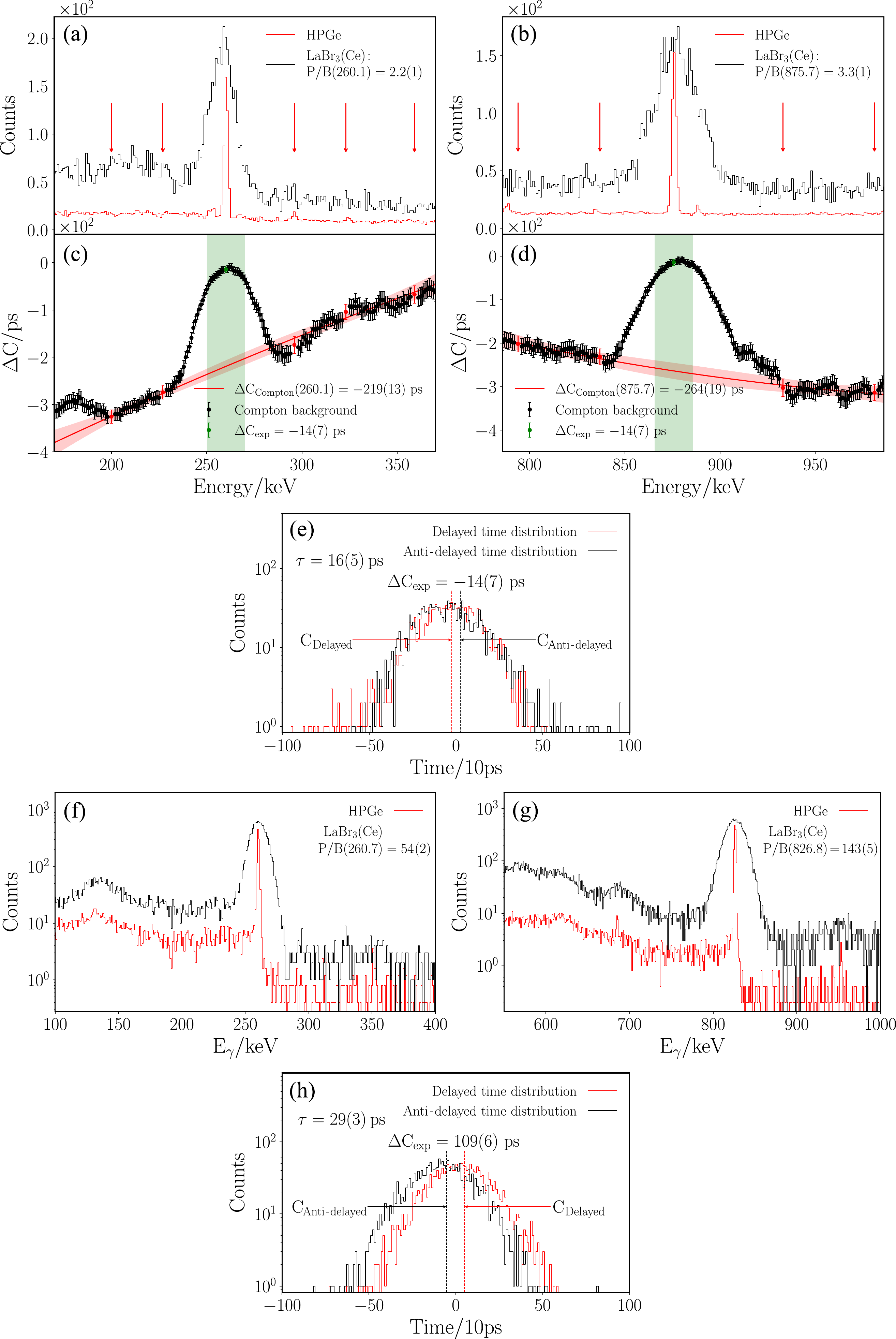}
	\caption{(a) Energy spectrum of the 260.1\,keV feeding transition of the $5/2_1^-$ state in $^{93}$Y using a LaBr$_3$(Ce) gate on the $5/2_1^- \rightarrow 1/2_{\text{g.s.}}^-$ (875.7\,keV) transition. (b) Energy spectrum of the 875.7\,keV decaying transition of the $5/2_1^-$ state in $^{93}$Y using a LaBr$_3$(Ce) gate on the $(3/2^+, 5/2^-) \rightarrow 5/2_1^-$ (260.1\,keV) transition. (c,d) Interpolated Compton background time response of the corresponding transition. Data points in red [also labelled in the energy spectrum (a,b)] are used for the fitted curve. Note that the scan over the entire background was performed in 1\,keV steps using 20\,keV gate widths. Therefore, only data points separated by at least 20 steps were used for the fit to avoid overlap between adjacent gates. The green bar represents the width of all used gates. (e) Delayed and anti-delayed time distributions for the $5/2_1^-$ state using LaBr$_3$(Ce) gates on the transitions mentioned above. \\
    Energy spectra of the feeding (f) and decaying (g) transitions of the $5/2_1^-$ state in $^{95}$Y either gating on the $5/2_1^- \rightarrow 1/2_{\text{g.s.}}^-$ decaying transition (826.8\,keV) or on the $9/2^+_1 \rightarrow 5/2_1^-$ feeding transition (260.7\,keV), and using an additional gate on the ionisation chamber (Note the logarithmic scale). Due to the cut of nearly all background and contamination except for transitions underneath isomeric states (e.g. $^{95m}$Y $(9/2_1^+)$), there is only one peak left in the spectra. (h) Resulting delayed and anti-delayed time distributions for the $5/2_1^-$ state in $^{95}$Y. \label{Results}}
\end{figure*}

\noindent The first $5/2^-$ state in $^{95}$Y is located at an excitation energy of 826.9\,keV. Its lifetime was determined using LaBr$_3$(Ce) gates on the $9/2_1^+ \rightarrow 5/2_1^- \rightarrow 1/2^-_{\text{g.s.}}$ (260.7\,keV$-$826.8\,keV) cascade. In addition, gating on the $A=95$ peak of the $\Delta E$ signal originating from the $Z$-dependent energy loss in the gas measured by the IC allows the cut of all other contaminants having different masses but similar $A/q$ ratios. Furthermore, owing to the flight time ($\approx$ 1.7\,$\mu$s) of the fission fragments from production to the detection setup, the gate on the IC signal suppresses nearly all short-lived contaminant contributions originating from $\beta$ decay, which is predominantly populated after implantation of the nuclei into the aluminium foil. The only remaining contributions arise from transitions below isomeric states with half-lives on the order of several tens of microseconds [e.g. $^{95m}$Y $(9/2_1^+)$, see Fig.~\ref{full_proj}]. As a result, an exceptionally clean spectrum is obtained, characterized by negligible background and a single remaining peak corresponding to either the feeding or decaying transition, depending on the selected initial LaBr$_3$(Ce) gate (see Figs.~\ref{Results}~(f,g)). This leads to extremely large peak\nobreakdash-to\nobreakdash-background ratios of 54 for the 260.7\,keV transition and 143 for the 826.8\,keV transition. Consequently, no background correction for the time-correlated Compton component is required. The (anti\nobreakdash-)delayed time distributions obtained after applying gates on the transitions mentioned before are shown in Fig.~\ref{Results}~(h). The deduced lifetime is given as $\tau_{{5/2}_1^-} = 29(3)$\,ps.
\renewcommand{\arraystretch}{1.1}
\begin{table*}[t]
	\begin{ruledtabular}
    \caption{Final results of the measured lifetimes in the odd-$A$ yttrium chain ($^{93-99}$Y) where $P/B$ refers to the peak-to-background ratios of the full-energy peaks. The "IC$/$HPGe" column indicates whether an additional gate was applied for further reduction of contaminants. Final adopted values used for further calculations are presented in bold.}
    	\begin{tabular}{ccccccccc}
			\multirow{2}{*}{Nucleus} & State & E$_{\text{state}}$ & Cascade & IC$/$HPGe & $P/B$ & $\tau_{\text{exp}}$ & $\tau_{\text{adopted}}$ & $\tau_{\text{literature}}$ \\ 
			& $J^{\pi}$ & keV & keV $-$ keV & keV & feeder $-$ decay & ps & ps & ps \\ \hline
			\rule{0pt}{3.0ex}& \multirow{2}{*}{$(3/2_1)^-$} & \multirow{2}{*}{590.2} & 168.5 $-$ 590.2 &   & 3.9(1) $-$ 9.0(1) & 3(2) &  \multirow{2}{*}{\textbf{$\leq$\,5}} & \multirow{2}{*}{$\leq\,2\times\,10^{3}$} \multirow{2}{*}{\cite{Herzog1972}} \\ 
			& & & 710.3 $-$ 590.2 &  & 7.7(2) $-$ 12.5(3) & -2(3) & & \\ 
            \rule{0pt}{3.5ex}$^{93}$Y & $5/2_1^-$ & 875.9 & 260.1 $-$ 875.7 &   & 2.2(1) $-$ 3.3(1) & 16(5) & \textbf{16(5)} &  \\ 
            \rule{0pt}{3.5ex}& \multirow{2}{*}{$(3/2^+, 5/2^-)$} & \multirow{2}{*}{1300.5} & 346.5 $-$ 710.3 &   & 0.8(1) $-$ 2.3(1) & 20(5) &  \multirow{2}{*}{\textbf{~~~15(3)\,$^{\rm a}$}} &  \\ 
			& & & 1269.5 $-$ 710.3 &  & 4.3(2) $-$ 6.3(2) & 12(4) & &  \\[0.7ex] \hline
            \rule{0pt}{3.0ex}&  &   & 945.0 $-$ 685.7 &   & 1.0(1) $-$ 3.3(1) & -6(3) &  & \\ 
			& $3/2^-_1$ & 685.9 & 1277.4 $-$ 685.7 &  & 0.8(1) $-$ 5.1(1) & 6(3) & \textbf{$\leq$\,9} &  \\ 
            \multirow{3}{*}{$^{95}$Y}&  &  & 1360.9 $-$ 685.7 &   & 0.3(1) $-$ 3.4(1) & 1(8) &  & \\ 
            \rule{0pt}{3.5ex}& \multirow{2}{*}{$5/2_1^-$} & \multirow{2}{*}{826.9} & 260.7 $-$ 826.8 & IC & 54(2) $-$ 143(5) & 29(3) & \multirow{2}{*}{\textbf{~~~27(2)}} &  \\ 
			& & & 260.7 $-$ 826.8 &  & 32.9(3) $-$ 51.9(4) & 27(2) & & \\
            \rule{0pt}{3.5ex}& $(5/2_2^-)$ & 1630.9 & 576.6 $-$ 945.0 &  & 0.7(1) $-$ 1.2(1) & ~~~12(3)\,$^{\rm b}$ & \textbf{$\leq$\,15} &  \\[0.7ex] \hline
            \rule{0pt}{3.0ex}& \multirow{2}{*}{$(3/2^-, 5/2^-)$} & \multirow{2}{*}{953.8} & 365.8 $-$ 953.8 &  & 1.4(1) $-$ 1.6(1) & ~~~9(3)\,$^{\rm b}$ & \multirow{2}{*}{\textbf{$\leq$\,11}} & \multirow{2}{*}{$\leq$\,6} \multirow{2}{*}{\cite{PhysRevC.41.1115}} \\ 
			& & & 1258.0 $-$ 953.8 &  & 1.5(1) $-$ 4.3(1) & ~~~8(3)\,$^{\rm b}$ & &  \\
            \rule{0pt}{3.5ex}& \multirow{2}{*}{$(5/2_1^+)$} & \multirow{2}{*}{1319.5} & 801.6 $-$ 652.2 &   & 1.9(1) $-$ 1.6(1) & 22(5) & \multirow{2}{*}{\textbf{~~~24(4)\,$^{\rm a}$}} & \multirow{2}{*}{18(7)} \multirow{2}{*}{\cite{PhysRevC.41.1115}} \\ 
			& & & 892.2 $-$ 652.2 &  & 1.2(1) $-$ 1.5(1) & 27(6) & &  \\
            \rule{0pt}{3.5ex}$^{97}$Y & ? & 1336.0 & 321.3 $-$ 668.5 &  & 0.7(1) $-$ 1.1(1) & ~~~6(4)\,$^{\rm b}$ & \textbf{$\leq$\,10} & $\leq\,4\times\,10^{3}$ \cite{Lhersonneau1986}\\ 
            \rule{0pt}{3.5ex}& $(13/2_1^+)$ & 1657.4 & 911.4 $-$ 321.3 &  & 1.1(1) $-$ 0.7(1) & 15(7) & \textbf{15(7)} & $\leq\,3\times\,10^{3}$ \cite{Lhersonneau1986}\\ 
			\rule{0pt}{3.5ex}& $1/2^+, 3/2^+$ & 1904.9 & 307.1 $-$ 1905.0 &  & 4.7(1) $-$ 3.8(1) & ~~~6(4)\,$^{\rm c}$ & \textbf{$\leq$\,10} & $\leq$\,3.3 \cite{PhysRevC.41.1115}\\ 
            \rule{0pt}{3.5ex}& $(17/2_1^+)$ & 2568.8 & 791.7 $-$ 911.4 &  & 1.7(1) $-$ 2.0(1) & ~~~9(4)\,$^{\rm b}$ & \textbf{$\leq$\,13} & $\leq \,3\times\,10^{3}$ \cite{Lhersonneau1986}\\[0.7ex] \hline
            \rule{0pt}{3.0ex}& \multirow{2}{*}{$(7/2_1^+)$} & \multirow{2}{*}{125.1} & \multirow{2}{*}{158.6 $-$ 125.1} & ~~198.5, 223.9\,\multirow{2}{*}{$^{\rm d}$} & \multirow{2}{*}{2.2(1) $-$ 5.0(1)} & \multirow{2}{*}{64(10)} & \multirow{2}{*}{\textbf{64(10)}} & 82(14) \cite{Rudigier},\\
            & & & & 214.0, 272.9 & & & & 68(9) \cite{PhysRevC.41.1141,WOHN1990141}\\
            \rule{0pt}{3.5ex}\multirow{4}{*}{$^{99}$Y} & \multirow{2}{*}{$(9/2_1^+)$} & \multirow{2}{*}{283.7} & \multirow{2}{*}{198.5 $-$ 158.6} & \multirow{2}{*}{~~~125.1, 223.9\,$^{\rm d}$} & \multirow{2}{*}{2.6(1) $-$ 2.6(1)} & \multirow{2}{*}{28(9)} & \multirow{2}{*}{\textbf{28(9)}} & 29(6) \cite{Rudigier}\\ 
            & & & & & & & & 26.9(19) \cite{PhysRevC.95.034302} \\
            \rule{0pt}{3.5ex}& \multirow{2}{*}{$(11/2_1^+)$} & \multirow{2}{*}{482.2} & \multirow{2}{*}{223.9 $-$ 198.5} & \multirow{2}{*}{~~~125.1, 158.6\,$^{\rm d}$} & \multirow{2}{*}{1.9(1) $-$ 2.7(1)} & \multirow{2}{*}{2(12)} & \multirow{2}{*}{\textbf{$\leq$\,14}} & $\leq$\,22 \cite{Rudigier}\\ 
            & & & & & & & & 6.2(6) \cite{PhysRevC.95.034302}\\
            \rule{0pt}{3.5ex}& $(13/2_2^+)$ & 1868.7 & 272.9 $-$ 214.0 & 125.1 & 3.3(1) $-$ 3.0(1) & ~~~29(6)\,$^{\rm e}$ & & 33(9) \cite{Rudigier}\\ [0.7ex]
		\end{tabular}
	  \label{tab:alldata}
    \end{ruledtabular}
    \begin{itemize}
  	\item[${}^{\rm a}$] Weighted average.
	\item[${}^{\rm b}$] Due to insufficient peak-to-background ratios and$/$or low statistics in data points used for the background correction, upper limits were adopted as final values.
	\item[${}^{\rm c}$] Due to the extrapolation of the time walk for the 1905.0\,keV transition an upper limit was adopted as final value.
	\item[${}^{\rm d}$] Matrices with multiple HPGe gates have been added to increase the statistic.
	\item[${}^{\rm e}$] The value was not adopted because the 272.9\,keV transition exhibits slight contamination of the 269.6\,keV transition.
    \end{itemize}
\end{table*}
\subsection{Experimental results}
\noindent Comparison of the lifetimes measured in this work with literature values shows excellent agreement. All measured lifetimes are consistent within the 1$\sigma$ uncertainties. In one case, the uncertainty was significantly reduced. Furthermore, the lifetimes of five states were determined whose lifetimes were previously unknown. In addition, lifetimes of four states previously reported to be smaller than 2$-$4\,ns are now found to lie in the low ps range, reflecting the improved timing capabilities of LaBr$_3$(Ce) detectors compared to earlier measurements employing lithium\nobreakdash-doped and standard germanium detectors. \\
The first $(3/2)^-$ state in $^{93}$Y was investigated using two independent cascades. In this case, the limits of the fast-timing method are approached, as both extracted lifetimes are consistent with zero; one even yields a slightly negative value despite sufficient peak-to-background ratios. Consequently, an upper limit of 5\,ps is adopted. The corresponding literature value, measured in 1972 using lithium-doped germanium detectors~\cite{Herzog1972}, provided only an upper limit of 2\,ns due to limited time resolution. The $5/2_1^-$ and $(3/2^+, 5/2^-)$ states were measured for the first time. For the $(3/2^+, 5/2^-)$ state, two cascades were analyzed independently, yielding consistent results within 1$\sigma$. Therefore, a weighted average was adopted as the final value. \\
In $^{95}$Y, all lifetimes were determined for the first time. Three cascades were available for the first $3/2^-$ state; however, due to limitations similar to those encountered for the first $(3/2)^-$ state in $^{93}$Y, as well as poorer peak\nobreakdash-to\nobreakdash-background ratios, only an upper limit could be deduced. The exceptional precision achieved for the first $5/2^-$ state, as discussed in Sec.~\ref{IC_state}, arises from the analysis of both twofold [LaBr$_3$(Ce)$-$LaBr$_3$(Ce)] and threefold [IC$-$LaBr$_3$(Ce)$-$LaBr$_3$(Ce)] coincidences. The threefold coincidences provide a substantial background reduction, while the twofold data offer very good statistical quality. As a result, an exceptionally small uncertainty is obtained. For the $(5/2_2^-)$ state, only an upper limit could be assigned due to low peak\nobreakdash-to\nobreakdash-background ratios and sensitivity of the extracted lifetime to the choice of background correction parameters. \\
In $^{97}$Y, the states $(3/2^-, 5/2^-)$ at 953.8\,keV, $(5/2_1^+)$, and $1/2^+, 3/2^+$ at 1904.9\,keV were previously investigated in 1990~\cite{PhysRevC.41.1115}. The experimental setup consisted of barium fluoride (BaF$_2$) detectors instead of LaBr$_3$(Ce) and an additional plastic detector enabled selection of specific cascades following $\beta$ decay. The present results confirm the previously reported lifetimes, with a notable reduction in the uncertainty for the $(5/2_1^+)$ state. Here, two different cascades are accessible leading to an weighted average as final adopted value. In this work, the lifetime of the $1/2^+, 3/2^+$ state at 1904.9\,keV is reported as an upper limit due to the required extrapolation of the time walk for the 1905.0\,keV transition. Additional states, namely "?" at 1336.0\,keV, $(13/2_1^+)$, and $(17/2_1^+)$, were studied in 1985~\cite{Lhersonneau1986} using germanium-based detector systems, yielding upper limits of 3$-$4\,ns. The improved time resolution of LaBr$_3$(Ce) detectors reduces these values to the picosecond range. For the "?" and $(17/2_1^+)$ states, upper limits are adopted due to low peak-to-background ratios and systematic uncertainties associated with background correction, similar to those observed for the $(5/2_2^-)$ state in $^{95}$Y. Both peaks used for the determination of the lifetime of the $(13/2_1^+)$ state have also low peak-to-background ratios. However, the statistics in background data points are quite sufficient. Therefore, its lifetime is associated with a relatively large uncertainty of approximately 50\,\%. \\
All lifetimes in $^{99}$Y were extracted using threefold coincidences. In addition, matrices gated with different HPGe conditions were summed to improve statistical quality. A previous experiment performed at LOHENGRIN in 2012~\cite{Rudigier} employed a nearly identical setup. Comparison with these results shows a reduction in uncertainties for the $(7/2_1^+)$ and $(13/2_2^+)$ states in the present work. Although Ref.~\cite{Rudigier} also utilized threefold coincidences involving the IC to suppress background, the present approach—combining multiple HPGe-gated threefold datasets—further enhances peak-to-background ratios of the FEPs, likely accounting for the improved precision. Furthermore, the lifetime for the $(7/2_1^+)$ state was previously measured at the Brookhaven National Laboratory using the TRISTAN mass separator in 1990~\cite{PhysRevC.41.1141,WOHN1990141}. The setup, consisting of a germanium, BaF$_2$, and a plastic detector, was similar to that used in Ref.~\cite{PhysRevC.41.1115}, yielding consistent results with the present work. Literature values for the $(9/2_1^+)$ and $(11/2_1^+)$ states were obtained from a RDDS measurement at GANIL in 2016~\cite{PhysRevC.95.034302}. Given the high sensitivity of RDDS techniques in the picosecond regime, no significant improvement in precision is achieved in the present work. Nevertheless, the measured lifetimes are consistent with the literature values and agree within the 1$\sigma$ interval, despite the exceptionally small uncertainties reported in the RDDS study.
\subsection{Isomeric yields}
\noindent Intensities of $\gamma$ rays that are not perturbed by background in the ungated single germanium spectra were used to determine the relative intensity of IT decays from the $(17/2^+_2)$ isomer in $^{99}$Y and put into relation to $\gamma$ rays from $\beta\gamma$ decays of $^{99\text{g}}$Y and $^{99}$Sr, respectively. \\
The isomeric ratio defined as the ratio of independent yields [$^{99\text{m}}$Y] / ([$^{99\text{g}}$Y] + [$^{99\text{m}}$Y]) was determined from the relative intensities (cumulative yields) [$^{99\text{m}}$Y] $/$ ([$^{99\text{g}}$Y] $-$ [$^{99}$Sr]) and corrected for in-flight decay losses between target and focal plane. Thus, an isomeric ratio of 2.8(6)\% has been found for $^{235}$U(n$_{\text{th}}$,f). We note that this ratio has been determined only for one given ionic charge and kinetic energy setting, but the result matches what had been measured in neutron-induced fission of $^{239}$Pu and $^{241}$Pu, respectively: IR = 2.5$-$3\% at peak energy~\cite{Chebboubi2025LOHENGRIN}. \\
In the same way the isomeric yield of $^{95\text{m}}$Y has been determined to 67(14)\%. As expected for a medium-spin isomer in the yrast cascade, its fission population is significantly higher compared to $^{99\text{m}}$Y. We note that $^{95}$Y is mainly indirectly fed by $\beta$ decay of $^{95}$Sr, i.e. the independent $^{95}$Y production has to be deduced from the difference of $^{95}$Y decays (the latter deduced with the most recent $\gamma$-ray intensities from Ref.~\cite{PhysRevC.108.014324}) and $^{95}$Sr decays. Therefore, the deduced IR is very sensitive to changes in the absolute $\gamma$ intensities of these decays.
\section{THEORETICAL FRAMEWORK}\label{sec:framework}
\noindent The development of new theoretical models is essential to enhance our understanding of nucleon behavior in this mass region, especially for odd-$A$ nuclei. In these nuclei, the complexity of modeling low-lying energy levels is greater than in even-even nuclei due to the additional interactions between the bosonic core and a single fermion. As a result, testing and comparing new models against experimental data provides valuable insights into the nuclear structure and collective nucleon dynamics. In this study, we perform calculations using the interacting boson-fermion model (IBFM) framework~\cite{IachelloVanIsackerBook}, initially proposed by Iachello and Scholten in Ref.~\cite{PhysRevLett.43.679}. We employ the interacting boson-fermion model with configuration mixing (IBFM-CM)~\cite{Gavrielov2022c,Gavrielov2023a}, in which the odd-$A$ nucleus is treated as a system of monopole ($s$) and quadrupole ($d$) bosons representing valence nucleon pairs, coupled to a single unpaired proton. The model Hamiltonian takes the form:
\begin{equation}
  \hat H = \hat H_b + \hat H_f + \hat V_{bf},
\end{equation}
\noindent where $\hat H_b$ is the IBM-CM boson-core Hamiltonian, $\hat H_f$ is the fermion single-particle Hamiltonian, and $\hat V_{bf}$ is the boson\nobreakdash-fermion (BF) interaction~\cite{IachelloVanIsackerBook}. Two configurations (normal $A$ and intruder $B$) are mixed via off-diagonal terms in both $\hat H_b$ and $\hat V_{bf}$. We compare these results with experimental $E2$ and $M1$ transition rates for odd-$A$ $^{93-99}$Y nuclei. \\
There are different approaches considering IBFM models. In contrast to other IBFM models, the IBFM-CM does not distinguish between protons and neutrons but includes shape coexistence, and the results are obtained by fitting experimental data to the model. 
\renewcommand{\arraystretch}{1.1}
\begin{table*}
    \begin{ruledtabular}
    \caption{Comparison of $E2$ and $M1$ transition probabilities for negative-parity states in $^{93}$Y and $^{95}$Y, for negative- and positive-parity states in $^{97}$Y, and for positive-parity states in $^{99}$Y. Multipolarities and $B_{\text{exp}}$ values marked with an asterisk are extracted under the assumption of pure $M1$ or $E2$ character. For states without upper limits in the adopted values and with available theoretical calculations, mixing ratios are assumed to reproduce the calculations (denoted by index b). For transitions involving states with uncertain spin$/$parity assignments (e.g., $(3/2^+, 5/2^-)$ in $^{93}$Y), alternative multipolarities may be possible. The present IBFM-CM calculations are restricted to $E2$ and $M1$ transitions; only these are listed, except for well-established spin$/$parity assignments implying different multipolarities (e.g., $E1$, $M2$). A dash ($-$) indicates that no calculated value is available. For $B(M1)$: $1\,\text{W.u.} \approx 1.8\,\mu_N^2$. For $B(E2)$: $1\,\text{W.u.} = 25.0, 25.8, 26.5$, and $27.2\,e^2\mathrm{fm}^4$ for $^{93}$Y, $^{95}$Y, $^{97}$Y, and $^{99}$Y respectively. Calculated moments for $^{99}$Y: $\mu(5/2^+_1) = 3.23\,\mu_N, Q(5/2^+_1) = +1.34\;e$b; $\mu(9/2^+_1) = 4.91\,\mu_N, Q(9/2^+_1) = -0.43\;e$b. \label{tab:trans-93-95-neg}}
        \begin{tabular}{ccccccc}
            \multirow{2}{*}{Nucleus} & Transition & E$_{\gamma}$ & Multipolarity & \multirow{2}{*}{$\delta$} & $B_{\text{exp}}$($\sigma\lambda;J^{\pi_i}_i \rightarrow J^{\pi_f}_f$) & $B_{\text{theo}}$($\sigma\lambda;J^{\pi_i}_i \rightarrow J^{\pi_f}_f$)  \\ 
			& $J^{\pi_i}_i \rightarrow J^{\pi_f}_f$ & keV & $\sigma\lambda$ & & W.u. & W.u. \\ [0.7ex]\hline
            \rule{0pt}{3.0ex}\multirow{11}{*}{$^{93}$Y} & \multirow{2}{*}{$(3/2_1)^-\to 1/2^-_{\rm g.s.}$} & \multirow{2}{*}{590.2} & $M1^*$ & \multirow{2}{*}{<\,1.4\,$^{\rm a}$~\cite{PhysRevC.10.2526}} & ${\geq\,0.031}^*$, ${\geq\,0.01}\,^{\rm a}$ & 0.662 \\
            & & & $E2^*$ & & ${\geq\,90}^*$, ${\geq\,60}\,^{\rm a}$ & 6.35 \\ 
            \rule{0pt}{3.5ex}& \multirow{2}{*}{$5/2^-_1 \to (3/2_1)^-$} & \multirow{2}{*}{285.7} & $M1^*$ & \multirow{2}{*}{$0.341$\,$^{\rm b}$} & ${9^{+4}_{-2}\times10^{-4}}^*$, ${8^{+4}_{-2}\times10^{-4}}\,^{\rm b}$ & $2\times10^{-6}$ \\
            & & & $E2^*$ & & ${12^{+6}_{-3}}^*$, ${1.3^{+6}_{-3}}\,^{\rm b}$ & 1.25 \\
            & $5/2^-_1\to 1/2^-_{\rm g.s.}$ & 875.7 & $E2$ & & ${3.9^{+18}_{-9}}$ & 5.74 \\ 
            \rule{0pt}{3.5ex}& \multirow{2}{*}{$(3/2^+\!,5/2^-)\,^{\rm c}\to 5/2^-_1$} & \multirow{2}{*}{424.7} & $M1^*$ & & ${3.2^{+9}_{-7}\times10^{-4}}^*$ & $-$ \\
            & & & $E2^*$ & & ${1.8^{+5}_{-4}}^*$ & $-$ \\
            & $(3/2^+\!,5/2^-)\,^{\rm c}\to (9/2_1)^+$ & 541.9 & $-$ & & $-$ & $-$  \\
            & \multirow{2}{*}{$(3/2^+\!,5/2^-)\,^{\rm c}\to (3/2_1)^-$} & \multirow{2}{*}{710.3} & $M1^*$ & & ${5.6^{+14}_{-9}\times10^{-3}}^*$ & $-$ \\
            & & & $E2^*$ & & ${12^{+3}_{-2}}^*$ & $-$ \\ [0.7ex]\hline
            \rule{0pt}{3.0ex}\multirow{9}{*}{$^{95}$Y} & \multirow{2}{*}{$3/2^-_1\to 1/2^-_{\rm g.s.}$} & \multirow{2}{*}{685.7} & $M1^*$ & & ${\geq\,0.011}^*$ & 0.661 \\
            & & & $E2^*$ & & ${\geq\,23}^*$ & 6.65 \\ 
            \rule{0pt}{3.5ex}& \multirow{2}{*}{$5/2^-_1\to 3/2^-_1$} & \multirow{2}{*}{141.0} & $M1^*$ & \multirow{2}{*}{$0.112\,^{\rm b}$} & ${2.1^{+4}_{-4}\times10^{-3}}^*$, ${2.1^{+4}_{-4}\times10^{-3}}\,^{\rm b}$ & $9\times10^{-6}$ \\
            & & & $E2^*$ & & ${103^{+22}_{-21}}^*$, ${1.3^{+3}_{-3}}\,^{\rm b}$ & 1.27 \\
            & $5/2^-_1\to 1/2^-_{\rm g.s.}$ & 826.8 & $E2$ & & ${2.9^{+2}_{-2}}$ & 6.12 \\ 
            \rule{0pt}{3.5ex}& $(5/2^-_2)\to 9/2^+_1$ & 543.4 & $M2$ & & ${\geq\,40}$ & ~~~$-\,^{\rm d}$ \\
            & \multirow{2}{*}{$(5/2^-_2)\to 3/2^-_1$} & \multirow{2}{*}{945.0} & $M1^*$ & & ${\geq\,2.5\times10^{-3}}^*$ & 0.012 \\
            & & & $E2^*$ & & ${\geq\,2.8}^*$ & 6.1 \\[0.7ex]\hline
            \rule{0pt}{3.0ex}& \multirow{2}{*}{$(3/2^-\!,5/2^-)\to (1/2^-_{\rm g.s.})$} & \multirow{2}{*}{953.8} & $M1^*$ & & ${\geq\,3.3\times10^{-3}}^*$ & 0.614 \\
            & & & $E2^*$ & & ${\geq\,3.5}^*$ & 4.65 \\ 
            \rule{0pt}{3.5ex}\multirow{17}{*}{$^{97}$Y} & \multirow{2}{*}{$(5/2^+_1)\to(3/2^-\!,5/2^-)$} & \multirow{2}{*}{365.8} & $E1^*$ & & ${9^{+2}_{-2}\times10^{-5}}^*$ & ~~~$-\,^{\rm d}$ \\
            & & & $M2^*$ & & ${3.1^{+7}_{-5}\times10^{3}}^*$ & ~~~$-\,^{\rm d}$ \\
            & \multirow{2}{*}{$(5/2^+_1)\to\,{1/2, 3/2}\,^{\rm e}$} & \multirow{2}{*}{622.5} & $M1^*$ & & ${1.7^{+7}_{-6}\times10^{-4}}^*$ & $-$ \\
            & & & $E2^*$ & & ${0.4^{+2}_{-2}}^*$ & $-$  \\
            & $(5/2^+_1)\to (9/2_1)^+$ & 652.2 & $E2$ & & ${8^{+2}_{-1}}$ & 4.41 \\
            \rule{0pt}{3.5ex}& \multirow{2}{*}{$?\,^{\rm f}$ $\to (9/2_1)^+$} & \multirow{2}{*}{668.5} & $M1^*$ & & ${\geq\,0.011}^*$ & 0.326 \\
            & & & $E2^*$ & & ${\geq\,23}^*$ & 31.43 \\ 
            \rule{0pt}{3.5ex}& \multirow{2}{*}{$(13/2^+_1)\to\,?\,^{\rm f}$} & \multirow{2}{*}{321.3} & $M1^*$ & \multirow{2}{*}{$0.104\,^{\rm b}$} & ${0.02^{+2}_{-1}}^*$, ${0.02^{+2}_{-1}}\,^{\rm b}$ & 0.109  \\
            & & & $E2^*$ & & ${195^{+160}_{-65}}^*$, ${2.1^{+17}_{-7}}\,^{\rm b}$ & 2.07 \\
            & $(13/2^+_1)\to (9/2_1)^+$ & 989.9 & $E2$ & & ${1.4^{+11}_{-5}}$ & 5.8 \\
            \rule{0pt}{3.5ex}& \multirow{2}{*}{$1/2^+, 3/2^+\,^{\rm g}\to\,1/2, 3/2$} & \multirow{2}{*}{165.8} & $M1^*$ & & ${\geq\,0.008}^*$ & $-$ \\
            & & & $E2^*$ & & ${\geq\,296}^*$ & $-$ \\
            & \multirow{2}{*}{$1/2^+, 3/2^+\,^{\rm g}\to\,(5/2^+, 7/2^+)$} & \multirow{2}{*}{477.1} & $M1^*$ & & ${\geq\,6.3\times10^{-4}}^*$ & (0.00335,\,$-$)  \\
            & & & $E2^*$ & & ${\geq\,2.7}^*$ & (9.23,\,22.45) \\
            & \multirow{2}{*}{$1/2^+, 3/2^+\,^{\rm g}\to\,(5/2_1^+)$} & \multirow{2}{*}{585.2} & $M1^*$ & & ${\geq\,2.7\times10^{-4}}^*$ & 0.182 \\
            & & & $E2^*$ & & ${\geq\,0.8}^*$ & 58.61 \\
            & $1/2^+, 3/2^+\,^{\rm g}\to\,(3/2^-, 5/2^-)$ & 951.0 & $E1^*$ & & ${\geq\,2.9\times10^{-6}}^*$ & ~~~$-\,^{\rm d}$ \\
            & $1/2^+, 3/2^+\,^{\rm g}\to\,(1/2^-_{\rm g.s.})$ & 1905.0 & $E1^*$ & & ${\geq\,5.5\times10^{-6}}^*$ & ~~~$-\,^{\rm d}$ \\[0.7ex]
        \end{tabular}
    \end{ruledtabular}
\end{table*}
\subsection{Boson core and BCS calculation}\label{sec:bcs}
\noindent The even-even boson core for the $^{91-101}$Y chain is provided by the $^{90-100}$Sr isotopes ($Z = 38$), for which an IBM-CM fit was performed separately~\cite{gavrielov2026intertwinedquantumphasetransitions}. The BF interaction parameters used here are consistent with that core. For the odd proton, single-quasiparticle energies $\epsilon_j$ and occupation probabilities $v_j^2$ are obtained from a BCS calculation in the $Z = 28-50$ shell, occupying the $\pi(2p_{3/2},\,1f_{5/2},\,2p_{1/2},\,1g_{9/2})$ orbits. The empirical single-particle energies are taken from Table~XI of Ref.~\cite{Barea2009}, the same set used for the niobium calculation in Refs.~\cite{Gavrielov2022c,Gavrielov2023a}. For yttrium ($Z = 39$, 11 valence protons), the pairing gap $\Delta_F = 0.89$\,MeV was adjusted to reproduce the empirical proton pairing gap inferred from the adjacent even-even strontium isotopes.
\subsection{Boson-fermion interaction}\label{sec:bfint}
\noindent The BF interaction $\hat V_{bf}$ contains monopole ($A_0$), quadrupole ($\Gamma_0$), and exchange ($\Lambda_0$) terms, whose strengths are obtained by fitting to the energy spectra of each isotope~\cite{gavrielov2026configurationcrossingshapeevolution}. For the positive-parity sector only the single-$j$ orbit $\pi(1g_{9/2})$ contributes; for the negative-parity sector the three orbits $\pi(2p_{1/2},\,2p_{3/2},\,1f_{5/2})$ contribute. As in Refs.~\cite{Gavrielov2022c,Gavrielov2023a}, the same BF strengths are used for both configurations $A$ and $B$. The positive- and negative-parity sectors are calculated independently, while energies of the negative-parity states are rigidly shifted to place the lowest calculated level at the experimental value.
\subsection{Electromagnetic transition operators}\label{sec:emops}
\noindent The $E2$ and $M1$ transition operators have the standard IBFM-CM form $\hat T(\sigma L) = \hat T_b(\sigma L) + \hat T_f(\sigma L)$ (see Ref.~\cite{Gavrielov2023a} and references therein). 

\noindent\textbf{$\bm{M1}$ operator.}\quad 
For the boson part, $g$-factor parameters are fitted to the even-even $^{90-100}$Sr data. For the fermion part, Schmidt values are used with a spin-quenching factor, yielding the effective spin $g$-factor $g_s~=~0.8744~\times~5.5857\,\mu_N~=~4.884\,\mu_N$ and orbital $g$\nobreakdash-factor $g_l = 1\,\mu_N$. The quenching factor was determined by fitting calculated magnetic moments to the available experimental values for the ground states of $^{91-101}$Y~\cite{gavrielov2026configurationcrossingshapeevolution}. 

\noindent\textbf{$\bm{E2}$ operator.}\quad
The boson effective charges $(e^{(A)},\,\chi^{(A)})$, $(e^{(B)},\,\chi^{(B)})$ are taken from the even-even strontium core fit. The fermion effective charge is 
\begin{align}
  f^{(2)}_{jj'} = -\frac{e_f}{\sqrt{5}}\langle r^2\rangle_N
                \langle j\|Y^{(2)}\|j'\rangle,
\end{align}
\begin{align}
  \langle r^2\rangle_N
    = 1.01219\,A^{1/3}\!\left(N+\tfrac{3}{2}\right)\;\text{fm}^2,
\end{align}
\noindent with $N = 4$ for the $g_{9/2}$ orbit and $N = 3$ for the $f,p$ orbits. The overall effective charge $e_f = -1.083\,e$ was determined by fitting the quadrupole moment of the $9/2^+$ isomeric state of $^{93}$Y~\cite{gavrielov2026configurationcrossingshapeevolution}.
\section{DISCUSSION}
\label{discuss}
\noindent All calculated $E2$ and $M1$ transition probabilities for the negative-parity states in $^{93}$Y and $^{95}$Y, for negative- and positive-parity states in $^{97}$Y, and for the positive-parity states in $^{99}$Y, together with available experimental values, are presented in Tab.~\ref{tab:trans-93-95-neg}. 
For states with uncertain spin$/$parity assignments, only $E2$ and $M1$ transitions are listed, although other multipolarities may be possible, as the present IBFM-CM calculations are restricted to these. In addition, multipolarities and $B_{\text{exp}}$ values marked with an asterisk are extracted under the assumption of pure $M1$ or $E2$ character, when mixing ratios are not known. Furthermore, for states whose final lifetimes were not adopted as upper limits and for transitions for which theoretical calculations are available but no experimental mixing ratios have been measured, mixing ratios capable to reproduce the calculations as good as possible are assumed.
\subsection{\texorpdfstring{Spherical region: $^{\textbf{91}-\textbf{97}}$Y}{}}
\label{sec:91-97y}
\renewcommand{\arraystretch}{1.1}
\begin{table*}[t]
\addtocounter{table}{-1}
    \begin{ruledtabular}
    \caption{(\textit{Continued.})}
        \begin{tabular}{ccccccc}
            \multirow{2}{*}{Nucleus} & Transition & E$_{\gamma}$ & Multipolarity & \multirow{2}{*}{$\delta$} & $B_{\text{exp}}$($\sigma\lambda;J^{\pi_i}_i \rightarrow J^{\pi_f}_f$) & $B_{\text{theo}}$($\sigma\lambda;J^{\pi_i}_i \rightarrow J^{\pi_f}_f$)  \\ 
			& $J^{\pi_i}_i \rightarrow J^{\pi_f}_f$ & keV & $\sigma\lambda$ & & W.u. & W.u. \\ [0.7ex]\hline
            \rule{0pt}{3.5ex}& \multirow{2}{*}{$(17/2^+_1)\to\,?$} & \multirow{2}{*}{94.0} & $M1^*$ & & ${\geq\,0.03}^*$ & $-$ \\
            & & & $E2^*$ & & ${\geq\,3.3\times10^{3}}^*$ & $-$ \\
            $^{97}$Y & \multirow{2}{*}{$(17/2^+_1)\to\,?$} & \multirow{2}{*}{452.7} & $M1^*$ & & ${\geq\,6\times10^{-4}}^*$ & $-$  \\
            & & & $E2^*$ & & ${\geq\,2.9}^*$ & $-$ \\
            & $(17/2^+_1)\to (13/2_1^+)$ & 911.4 & $E2$ & & ${\geq\,3.4}$  & 107.22 \\[0.7ex]\hline
            \rule{0pt}{3.0ex}\multirow{12}{*}{$^{99}$Y} & \multirow{2}{*}{$(7/2^+_1)\to (5/2^+_{\rm g.s.})$} & \multirow{2}{*}{125.1} & $M1^*$ & \multirow{2}{*}{0.118\,$^{\rm b}$} & ${0.23^{+4}_{-3}}^*$, ${0.23^{+4}_{-3}}\,^{\rm b}$ & 0.00235 \\
            & & & $E2^*$ & & ${1.0^{+2}_{-1}\times10^{4}}^*$, ${190^{+35}_{-26}}\,^{\rm b}$ & 190.26 \\
            \rule{0pt}{3.5ex}& \multirow{2}{*}{$(9/2^+_1)\to (7/2^+_1)$} & \multirow{2}{*}{158.6} & $M1$ & \multirow{2}{*}{0.17(5)~\cite{PhysRevC.95.034302}} & $0.24^{+11}_{-6}$ & 0.02505  \\
            & & & $E2$ & & $266^{+231}_{-139}$ & 159.66  \\
            & $(9/2^+_1)\to (5/2^+_{\rm g.s.})$ & 283.7 & $E2$ & & ${45^{+23}_{-13}}$ & 58  \\
            \rule{0pt}{3.5ex}& \multirow{2}{*}{$(11/2^+_1)\to (9/2^+_1)$} & \multirow{2}{*}{198.5} & $M1$ & \multirow{2}{*}{0.16(5)~\cite{PhysRevC.95.034302}} & $\geq\,0.2$ & 0.06951 \\
            & & & $E2$ & & $\geq\,67$ & 108.08 \\
            & $(11/2^+_1)\to (7/2^+_1)$ & 357.2 & $E2$ & & ${\geq\,49}$ & 92.45\\
            \rule{0pt}{3.5ex}& \multirow{2}{*}{{\color{gray} $(13/2^+_2)\to (11/2^+_2)$}} & \multirow{2}{*}{{\color{gray} 214.0}} & {\color{gray} $M1^*$} & \multirow{2}{*}{{\color{gray} 0.233\,$^{\rm b}$}} & {\color{gray} ${0.11^{+3}_{-2}}^*$, ${0.11^{+3}_{-2}}\,^{\rm b}$} & {\color{gray} 0.34271}  \\
            & & & {\color{gray} $E2^*$} & & {\color{gray} ${2.2^{+6}_{-4}\times10^{3}}^*$, ${117^{+30}_{-21}}\,^{\rm b}$} & {\color{gray} 117.01}  \\ [0.7ex]
        \end{tabular}
    \end{ruledtabular}
    \begin{itemize}
    \item[${}^{\rm a}$] $B_{\text{exp}}$ calculated using the maximum of the given mixing ratio.
    \item[${}^{\rm b}$] Assumed mixing ratios and resulting $B_{\text{exp}}$ values.
    \item[${}^{\rm c}$] The state at 1300.5\,keV is likely outside the IBFM-CM model space, corresponding to a proton coupling to the $2^+_\text{2-4}$ excitations of $^{92}$Sr that are not included in the boson description.
	\item[${}^{\rm d}$] Inter-parity ($E1$ or $M2$) rates are not calculated by the present IBFM-CM code.
    \item[${}^{\rm e}$] The calculation predicts no $1/2^+$ or $3/2^+$ state below the $5/2^+_1$. This transition may instead correspond to the $3/2^-$ state.
    \item[${}^{\rm f}$] The calculation column presents the results for an assumed $11/2^+_1$ state.
    \item[${}^{\rm g}$] Assuming the $(1/2^+, 3/2^+)$ is a $3/2^+$ in the calculation. The results in parentheses correspond to the experimental spin assignment in parentheses.
    \end{itemize}
    \label{tab:trans-99}
\end{table*}
\noindent For $N = 52-56$ ($^{91-95}$Y), the structure is governed by the normal $A$ configuration, with the adjacent $^{90-94}$Sr cores remaining nearly spherical. At $N = 58$ ($^{97}$Y), the calculated $9/2^+_1$ state remains predominantly normal, weakly deformed, while several other low-lying positive-parity states acquire dominant intruder-deformed components.
\subsubsection{Negative-parity states}\label{sec:neg-before}
\noindent The lowest three negative-parity states in $^{91-97}$Y are single quasiparticle (SqP) states of the $\pi(2p_{1/2})$, $\pi(2p_{3/2})$, and $\pi(1f_{5/2})$ orbits, each coupled to the $0^+_{1;A}$ of the adjacent strontium core, giving $J^\pi = 1/2^-, 3/2^-, 5/2^-$, respectively. Above the SqP, the coupling of these orbits to the $2^+_{1;A}$ (a spherical $n_d = 1$ state) of the strontium core generates multiplets. In this work, we only observe the $2^+_{1;A} \otimes \pi(2p_{1/2})$ doublet, which generates the $3/2^-$ and $5/2^-$ states. In $^{91,93}$Y, additional negative\nobreakdash-parity states observed experimentally likely arise from proton couplings to the $2^+_{2,3,4}$ states of the $^{90,92}$Sr cores. \\
The $3/2^-_1\to 1/2^-_{\rm g.s.}$ transition connects the $p_{3/2}$ and $p_{1/2}$ SqP states; the calculated $B(M1)\approx 0.66$\,W.u.\ is characteristic of this intra-shell transition. The experimental pure-$M1$ values are substantially smaller, with the $M1/E2$ mixing ratio playing a role in the extraction. The calculated result cannot be reproduced by varying the mixing ratio up to the literature value (<\,1.4). However, adopting a shorter lifetime ($\leq$\,1\,ps), consistent with the experimental upper limit of $\leq$\,5\,ps, resolves this discrepancy. For the $5/2^-_1\to 3/2^-_1$ transition, the calculation predicts essentially pure $E2$ ($B(M1)\approx 0$) with $B(E2)= 1.25$\,W.u for $^{93}$Y and 1.27\,W.u for $^{95}$Y. The experimental pure-$E2$ values of 12$^{+6}_{-3}$\,W.u.\ ($^{93}$Y) and 103$^{+22}_{-21}$\,W.u.\ ($^{95}$Y) are larger than calculated; the value of 103\,W.u.\ for $^{95}$Y appears large for a nearly spherical nucleus, and is expected to take a similar value as $^{93}$Y. Adopting mixing ratios of 0.341 for the extracted $B(E2)$ value in $^{93}$Y and 0.112 for $^{95}$Y reduces the corresponding former pure-$E2$ values to 1.3$^{+6}_{-3}$\,W.u.\ and 1.3$^{+3}_{-3}$\,W.u., respectively, yielding agreement with the calculations.
\subsubsection{Positive-parity states}\label{sec:pos-before}
\noindent The lowest positive-parity state in $^{91-97}$Y is $9/2^+_1$, originating from the weak coupling of the $\pi(1g_{9/2})$ proton to the $0^+_{1;A}$ of the adjacent strontium core. Above it, there is the coupling of $\pi(1g_{9/2})$ to the $2^+_{1;A}$ of the strontium core. This $2^+_{1;A}$ is a normal configuration-$A$ spherical state with a dominant U(5) $n_d=1$ component in its wave function. It generates a quintuplet of states such that $2^+_{1;A} \otimes \pi(1g_{9/2}) = 5/2,\,7/2,\,9/2,\, 11/2,\,13/2$. In the weak coupling regime, the $B(E2)$ transitions from the quintuplet members to the $9/2^+_1$ ground state are expected to be comparable to the $B(E2;\,2^+_1 \to 0^+_1)$ of the respective strontium core, as discussed for the analogous quintuplet in the niobium isotopes~\cite{Gavrielov2023a}. In the current $^{97}$Y calculation, however, configuration mixing substantially modifies this simple picture: the $9/2^+_1$ and $11/2^+_1$ states are 98.2\% and 98.9\% normal, whereas the $5/2^+_1$ and $13/2^+_1$ states are 93.4\% and 83.8\% intruder. Consequently, the $5/2^+_1 \to 9/2^+_1$ and $13/2^+_1 \to 9/2^+_1$ transitions occur predominantly between states with different dominant configurations. The corresponding calculated strengths are:
\begin{itemize}
    \item[] $B(E2;\,5/2^+_1\,\to\,9/2^+_1) = 4.41$\,W.u.,
    \item[] $B(E2;\,11/2^+_1\,\to\,9/2^+_1) = 31.43$\,W.u.,
    \item[] $B(E2;\,13/2^+_1\,\to\,9/2^+_1) = 5.8$\,W.u.;
\end{itemize}
their spread reflects the mixed configuration content rather than representing the direct weak-coupling image of the corresponding $^{96}$Sr-core transitions\footnote{For $^{91,93}$Y, the $^{90,92}$Sr cores carry additional $2^+_2,\,2^+_3,\,2^+_4$ states in the 1$-$2,MeV range, that are assumed to be related 1p-1h proton single-particle excitations, which are outside the IBM-CM model space. Similarly, the $4^+_{1;A}$ strontium states are assumed to be outside the IBM-CM model space (also for $^\text{94-96}$Sr, see \cite{Gavrielov2023a} for a similar case in Zr$-$Nb). The coupling of such states to the $\pi(1g_{9/2})$ produce extra positive-parity states outside the IBFM-CM model space. In contrast, $^{94}$Sr (core for $^{95}$Y) has no low-lying $2^+_\text{2-4}$ states.}. The calculated $B(E2;\,13/2^+_1 \to 9/2^+_1) = 5.8$\,W.u., while larger than the experimental value of $1.4^{+11}_{-5}$\,W.u., remains weaker than the $B(E2;\,2^+_1 \to 0^+_1) = 13(8)$\,W.u. transition strength of the $^{96}$Sr core. The calculated $B(M1;\,11/2^+_1\to 9/2^+_1) = 0.326$\,W.u.\ reproduces the $\geq$\,0.011 experimental value. Within the low-lying positive-parity states, the calculation gives: 
\begin{itemize}
    \item[] $B(M1;\,7/2^+_1 \to 9/2^+_1) = 0.412$\,W.u.,
    \item[] $B(M1;\,5/2^+_1 \to 7/2^+_1) = 0.217$\,W.u.,
    \item[] $B(E2;\,5/2^+_1 \to 7/2^+_1) = 16.31$\,W.u.,
\end{itemize}
(not shown in Tab.~\ref{tab:trans-93-95-neg}). For the $(13/2^+_1) \to 11/2^+_1$ transition, the calculation gives $B(E2) = 2.07$\,W.u. and $B(M1)~=~0.109$\,W.u. With an assumed mixing ratio of 0.104, the experimental $B(E2)$ value is reduced from the pure-$E2$ value of $195^{+160}_{-65}$ to $2.1^{+17}_{-7}$\,W.u., in agreement with the calculation, while the calculated $M1$ strength remains larger than the value inferred with this assumption.
\paragraph*{Comment on experimental levels.} The level at 1336.0\,keV is tentatively assigned as ($11/2^+$), or more conservatively as ($9/2^+, 11/2^+$). This assignment is supported by its placement in the isomeric-decay cascade of $^{97}$Y, between the established ($9/2^+$) level at 667.5\,keV and the ($13/2^+$) level at 1657.4\,keV, with the observed transitions listed as $D, E2$. In the interpretation of Ref.~\cite{Lhersonneau1996}, these states are associated with the $\pi g_{9/2} \otimes 2^+$ multiplet, for which the $9/2^+$ and $11/2^+$ members are expected below the $13/2^+$ state. The 1738.8\,keV level is more likely to have $J=3/2$ than $J=1/2$, since its decay to the 1428.1\,keV $(5/2^+,7/2^+)$ level is listed as dipole; a $1/2 \rightarrow 5/2$ transition would require $\Delta J=2$ and is therefore incompatible with a pure dipole assignment. Its parity is not fixed experimentally, but a tentative $(3/2^+)$ assignment is favored by the observed $\beta$ feeding from the $1/2^+$ parent and by the dipole feeding from the 1904.9\,keV $(1/2^+,3/2^+)$ level. The comparison between experimental and calculated energies of the positive-parity excited states in the odd-$A$ $^{91-101}$Y isotopes is presented in~\cite{gavrielov2026configurationcrossingshapeevolution}.
\subsection{\texorpdfstring{Deformed region: $^{\textbf{99}}$Y}{}}
\label{sec:after-cp}
\subsubsection{Onset of deformation and intruder dominance}
\noindent For $^{99}$Y ($N = 60$), the ground state becomes a $5/2_1^+$ state, which is the bandhead of a $K^\pi = 5/2^+$ rotational band in the intruder configuration, with members $5/2^+, 7/2^+, 9/2^+, 11/2^+, 13/2^+$,... At this point, the ground-state configuration crosses from the normal $A$ to the intruder $B$ configuration, marking the Type~II quantum phase transition identified in the adjacent even-even zirconium~\cite{PhysRevC.99.064324,Gavrielov2022c} and strontium~\cite{gavrielov2026configurationcrossingshapeevolution} chains, and odd niobium chain~\cite{Gavrielov2022c,Gavrielov2023a}. Simultaneously, the intruder configuration undergoes a gradual spherical-to-deformed shape evolution (Type~I QPT). The coexistence of both QPTs constitutes the intertwined QPTs (IQPTs) scenario established in the even-even strontium and zirconium chains and in the odd-$A$ niobium isotopes~\cite{Gavrielov2022c,Gavrielov2023a,Gavrielov2025} (See~\cite{Leviatan2025} for the geometrical interpretation).
\noindent The $\Delta J = 2$ transition $B(E2;\,9/2^+_1\to 5/2^+_{\rm g.s.}) = 58$\,W.u. is free of $M1/E2$ mixing; the experimental $45^{+23}_{-13}$\,W.u. is consistent with the calculation and supports the deformed description. The large in-band $\Delta J = 1$ calculated values ($B(E2) = 108-190$\,W.u.) reflect the collective nature of the intruder band. The corresponding experimental entries give excessively large pure-$E2$ values ($10^3-10^4$\,W.u.). Modifying the $M1/E2$ mixing ratio brings the $E2$ values down to a physically reasonable range, which could be consistent with the calculation. Assuming the mixing ratios that reduce the pure-$E2$ values to match the calculations yields values of 0.118 for the $(7/2^+_1)\to(5/2^+_{\text{g.s.}})$ transition and 0.233 for the $(13/2^+_2)\to(11/2^+_2)$ transition. Furthermore, adopting the mixing ratios reported in Ref.~\cite{PhysRevC.95.034302} also leads to excellent agreement with the calculations of the $B(E2)$ values.
\section{CONCLUSION} \label{sec:conclusion}
\noindent In this work, lifetimes of states in the odd-$A$ yttrium isotopes $^{93-99}$Y were investigated. All nuclei were measured in four different experiments at the LOHENGRIN recoil separator at the Institut Laue-Langevin. In $^{93}$Y, the states $(3/2_1)^-$, $5/2_1^-$, and the $(3/2^+,5/2^-)$ at 1300.5\,keV were investigated, while in $^{95}$Y the states $3/2_1^-$, $5/2_1^-$, and $(5/2_2^-)$ were measured. Here, all lifetimes were determined for the first time except for the $(3/2_1)^-$ state. In $^{97}$Y, the states $(3/2^-,5/2^-)$ at 953.8\,keV, $(5/2_1^+)$, the "?" at 1336.0\,keV, $(13/2_1^+)$, the $1/2^+,3/2^+$ at 1904.9\,keV, and $(17/2_1^+)$ were investigated. The extracted lifetimes of the $(3/2^-,5/2^-)$, $(5/2_1^+)$, and $1/2^+,3/2^+$ states are in good agreement with previously reported values~\cite{PhysRevC.41.1115}, while the uncertainty of the $(5/2_1^+)$ state was significantly reduced. The remaining three states in $^{97}$Y, as well as the $(3/2_1)^-$ state in $^{93}$Y, had previously only been constrained to values below 2$-$4\,ns~\cite{Herzog1972,Lhersonneau1986}. In the present work, these lifetimes were determined to lie in the low-ps range, demonstrating the improved timing capabilities of LaBr$_3$(Ce) detectors compared with earlier measurements employing lithium-doped and conventional germanium detectors. In $^{99}$Y, the states $(7/2_1^+)$, $(9/2_1^+)$, $(11/2_1^+)$, and $(13/2_2^+)$ were measured. All extracted lifetimes are in good agreement with literature values~\cite{Rudigier,PhysRevC.41.1141,WOHN1990141,PhysRevC.95.034302}. \\
Reduced transition probabilities, derived from the experimentally determined lifetimes, were compared with IBFM-CM calculations. Since most transitions are assumed to be of mixed multipolarity without known mixing ratios, pure $M1$ and $E2$ assumptions were adopted for the experimental estimates in most cases. In several instances, large discrepancies between experiment and theory are observed. In particular, the experimentally deduced $E2$ strengths are considerably larger than predicted, suggesting enhanced collective behavior, whereas the calculations favor predominantly single-particle transitions. The inclusion of realistic mixing ratios for transitions without experimentally established lower limits could reduce these discrepancies. In contrast, the comparison between experimental and theoretical transition probabilities in $^{99}$Y shows very good agreement for states with known mixing ratios, supporting the onset of collective behavior beyond the $N \approx 59$ deformation boundary. For states with uncertain spin$/$parity assignments, only $M1$ and $E2$ multipolarities were considered due to the limitations of the present calculations, although other multipolarities may also be possible. Where available, comparisons with additional theoretical approaches are also provided. Furthermore, the limited precision of transition probabilities reported only as lower limits restricts a more quantitative comparison between experiment and theory. These unknown properties add complexity and therefore make a clear comparison difficult for some states. \\
Future measurements with improved statistics and higher timing precision would help to further constrain the lifetimes and transition probabilities in the odd-$A$ yttrium isotopes. In particular, experimental determination of mixing ratios for transitions currently treated using assumed mixing ratios or pure $M1$ or $E2$ character would allow for a more reliable comparison with theoretical calculations and provide a more stringent test of the present model. Additional spin and parity assignments for presently uncertain states are also required to improve tests of theoretical models and clarify the underlying nuclear structure. Furthermore, extending lifetime measurements toward more neutron-rich yttrium isotopes beyond $N=60$ would provide valuable information on the evolution of collectivity and the onset of deformation in this mass region.
\section*{ACKNOWLEDGMENTS}
\noindent This work was supported by the Deutsche Forschungsgemeinschaft (DFG) under Grants No. JO 391/18-1 and JO 391/18-2. We also want to thank the ILL team of the nuclear reactor for its operation.
\appendix
\bibliography{ref.bib}
\end{document}